\documentclass[12pt]{article}
\usepackage{latexsym}
\usepackage{graphics}
\usepackage{color}
\newcommand{\gl}{\lambda}

\newcommand {\mD} {{\bf D}}

\newcommand {\mI} {{\bf I}}
\newcommand {\mM} {{\bf M}}
\newcommand {\mQ} {{\bf Q}}
\newcommand {\Tr} {\mbox{Tr}}
\newcommand {\Var} {\rm{Var}}

\begin{document}

\begin{titlepage}
\begin{flushright}
     {\bf UK/02-02}  \\
       Jan. 2002     \\
\end{flushright}

\begin{center}
{\large\bf Finite Density Algorithm in Lattice QCD -- \\ 
a Canonical Ensemble Approach}  
  
\vspace{1.5cm}

{Keh-Fei Liu\footnote{email: liu@pa.uky.edu \\
Invited talk at Nankai Symposium on Mathematical Physics, Tianjin, Oct. 2001 to 
be published in Journal of Modern Physics B.}}

\bigskip
\bigskip

{\it Dept. of Physics and Astronomy, University of Kentucky,
Lexington, KY 40506}
\end{center}

\vskip 12pt

\begin{abstract}
I will review the finite density algorithm for lattice QCD based on
finite chemical potential and summarize the associated difficulties.
I will propose a canonical ensemble approach which projects out
the finite baryon number sector from the fermion determinant. For this
algorithm to work, it requires an efficient method for calculating
the fermion determinant and a Monte Carlo algorithm which accommodates
unbiased estimate of the probability. I shall report on the
progress made along this direction with the Pad\'{e} - Z$_2$ estimator
of the determinant and its implementation in the newly developed 
Noisy Monte Carlo algorithm.
\end{abstract}

\vfill
\end{titlepage}

\section{Introduction}	

Fermions at finite density or finite chemical potential is
a subject of a wide range of interest. It is relevant to
condensed matter physics, such as the Hubbard model away from
half-filling. The research about nuclei and neutron stars at low and
high nucleon density is actively pursued in nuclear physics and
astrophysics. The subject of quark gluon plasma is important for understanding
the early universe and is being sought for in relativistic heavy-ion collisions
in the laboratories. Furthermore, speculation about color superconducting
phase has been proposed recently for quantum chromodynamics (QCD) at very high
quark density~\cite{arw98}.

Although there are models, e.g. chiral models and Nambu Jona-Lasinio model which
have been used to study QCD at finite quark density, the only way to
study QCD at finite density and temperature reliably and systematically is via
lattice gauge calculations. There have been extensive lattice calculations of QCD
at finite temperature~\cite{eji01}. On the contrary, the calculation at finite density
is hampered by the lack of a viable algorithm.

In this talk, I shall first review the difficulties associated with the
finite density algorithm with chemical potentials in Sec. 2. I will then
outline in Sec. 3 a proposal for a finite density algorithm in the canonical ensemble
which projects out the nonzero baryon number sector from the fermion determinant. In Sec. 4,
a newly developed Noisy Monte Carlo algorithm which admits unbiased estimate of
the probability is described. Its application to the fermion determinant is
outlined in Sec. 5. I will discuss an efficient way, the Pad\'{e}-Z$_2$ method, to 
estimate the $Tr \log$ of the fermion matrix in Sec. 6. The recent progress on the
implementation of the Kentucky Noisy Monte Carlo algorithm to dynamical fermions is 
presented in Sec. 7.  Finally, a summary is given in Sec. 8.

\section{Finite Chemical Potential}

The usual approach to the finite density in the Euclidean path-integral
formalism of lattice QCD is to consider the grand canonical ensemble
with the partition function
\begin{equation}   \label{ZGC}
Z_{GC}(\mu) = \sum_N Z_N e^{ - \mu N} = \int {\cal{D}} U det M[U,\mu] e^{-S_g[U]},
\end{equation}
where the fermion fields with fermion matrix $M$ has been integrated to give the
determinant. $U$ is the gauge link variable and $S_g$ is the gauge action.
The chemical potential is introduced to the quark action with the $e^{\mu a}$ factor
in the time-forward hopping term and $e^{ - \mu a}$ in the time-backward hopping
term. Here $a$ is the lattice spacing. However, this causes the fermion action
to be non-Hermitian, i.e. $\gamma_5 M \gamma_5 \neq M$. As a result, the
fermion determinant $det M[U]$ is complex and this leads to the infamous
sign problem.
%

There are several approaches to avoid the sign problem:

\subsection{Fugacity Expansion}

   It was proposed by the Glasgow group~\cite{bmk98} that
the sign problem can be circumvented based on the expansion
of the grand canonical partition function in powers of the
fugacity variable $e^{\mu/T}$,
\begin{equation}   \label{zgc}
Z_{GC} (\mu/T, T, V) = \sum_{B = - 3V}^{B = 3V} e^{\mu/T\, B} Z_B(T,V),
\end{equation}
where $Z_B$ is the canonical partition function for the baryon sector with
baryon number $B$. $Z_{GC}$ is calculated with reweighting of the
fermion determinant
\begin{equation}
Z_{GC} (\mu) = \langle \frac{det M[U,\mu]}{det M[U,0]} \rangle_{\mu = 0}.
\end{equation}
Since the reweighting is based on the gauge configuration with
$\mu = 0$, it avoids the sign problem. However, this does not work, except
perhaps at small $\mu$ or near the finite temperature phase transition.
We will dwell on this later in Sec. 3. This is caused by
the `overlap problem'~\cite{alf99} where the important samples of configurations
in the $\mu = 0$ simulation has exponentially small overlap with those
relevant for the finite density. As a result, the onset of baryon begins
at $\mu \sim m_{\pi}/2$ instead of the expected $M_N/3$ which
resembles the situation of the quenched approximation.

\subsection{Imaginary Chemical Potential}  \label{ICP}

   In this approach, the chemical potential is taking an
imaginary value $\mu = i \nu$. The fermion determinant is real
in this case and one can avoid the sign problem~\cite{dms90,wei87,akw00}. 
The partition function is
\begin{equation}
Z_{GC} (i\nu/T, T, V) = Tr\, e^{ - \hat{H}/T} e^{i \nu \hat{B}/T},
\end{equation}
which is periodic with respect to $\nu$ with a period of $2 \pi T$.
Comparing with Eq. (\ref{zgc}), one can in principle obtain canonical partition
function $Z_B$ from the Fourier transform
\begin{equation}
Z_B (T, V) = \frac{1}{2 \pi T} \int_0^{2 \pi T} d\nu Z_{GC}(i\nu/T, T, V)
e^{-i\nu \hat{B}/T}.
\end{equation}
In this approach, one needs to integrate over the whole range
of $\nu$ from $0$ to $2 \pi T$ after one obtains the Monte Carlo
configurations of $Z_{GC}(i\nu/T, T, V)$ at different $\nu$. In practice,
it is proposed to calculate the following ratio in the two-dimensional
Hubbard model~\cite{akw00},
\begin{equation}
\frac{Z_{GC}(i\nu/T, T, V)}{Z_{GC}(i\nu_0/T, T, V)}  =
\int {\cal{D}} \phi e^{-S_{bos}} det M(i\nu_0) \frac{det M(i\nu)}{det M(i\nu_0)},
\end{equation}
with a reference value $\nu_0$. Several patches each centered around
a different reference point $\nu_0$ are used to cover the range of
$\nu$. This was successful for the two-dimensional Hubbard model with
a $4^2 \times 10$ lattice up to B = 6 where the determinant was
calculated exactly. While this works for a small lattice in the
Hubbard model, it would not work for reasonably large lattices in QCD.
This is because the direct calculation of the determinant is
a $V^3$ (or $V^2$ for a sparse matrix) operation which is an impracticable
task for the quark matrix which is typically of the dimension 
$10^6 \times 10^6$. Any stochastic estimation of the
determinant will inevitably introduce systematic error. Furthermore, this
will also suffer from the `overlap' problem discussed above. Any Monte Carlo
simulation at a reference point $\nu_0$ will have exponentially small
overlap with those configurations important to a nonzero baryon density.

\subsection{Overlap Ensuring Multi-parameter Reweighting}

   To alleviate the sign problem with the real chemical potential and
the overlap problem due to reweighting, it is proposed~\cite{fk01} to
do the reweighting in the multiple parameter space. The generic 
partition function $Z_{GC}$ in Eq. (\ref{ZGC}) is parametrized
by a set of parameters $\alpha$, such as the chemical potential $\mu$,
the gauge coupling $\beta$, the quark mass $m_q$, etc.
The partition function can be written to facilitate reweighting
\begin{equation}
Z_{GC}(\alpha) = \int {\cal{D}} U det M[U,\alpha_0] e^{-S_g[U,\alpha_0]}
\{e^{-S_g[U,\alpha] + S_g[U,\alpha_0]} \frac{det M[U, \alpha]}{det M[U,\alpha_0]}\},
\end{equation}
where the Monte Carlo simulation is carried out with the $\alpha_0$
set of parameters and the terms in the curly bracket are treated as
observables. This is applied to study the end point in the T-$\mu$
phase diagram. In this case, the Monte Carlo simulation is carried
out where the parameters in $\alpha_0$ include $\mu = 0$ and $\beta_c$ which
corresponds to the phase transition at temperature $T_c$. The parameter
set $\alpha$ in the reweighted measure include  $mu \neq 0$ and
an adjusted $\beta$ in the gauge action. The new $\beta$ is determined 
from the Lee-Yang zeros so that one is following the transition line
in the T-$\mu$ plane and the large change in the determinant ratio
in the reweighting is compensated by the change in the gauge action to
ensure reasonable overlap. This is shown to work to locate the transition line
from $\mu =0$ and $ T = T_c$ down to the critical point on the
$4^4$ and $6^3 \times 4$ lattices with staggered fermions~\cite{fk01}.

   While the multi-parameter reweighting is successful near the
transition line, it is not clear how to extend it beyond this region,
particularly the $T = 0$ case where one wants to keep the $\beta$ and
quark mass fixed while changing the $\mu$. One still expects to face
the overlap problem in the latter case. For large volumes, 
calculating the determinant ratio will be subjected to the same practical 
difficulty as discussed in the previous section \ref{ICP}.

\section{Finite Baryon Density -- A Canonical Ensemble Approach}

   We would like to propose an algorithm to overcome the overlap problem
at zero temperature which is based on the canonical ensemble approach. To avoid 
the overlap problem, one needs to lock in a definite nonzero baryon sector so that
the exponentially large contamination from the zero-baryon sector is
excluded. To see this, we first note that the fermion
determinant is a superposition of multiple quark loops of all sizes
and shapes. This can be easily seen from the property of the determinant
\begin{equation} 
det M = e^{Tr \log M} = 1 + \sum_{n =1} \frac{(Tr \log M)^n}{n!}.
\end{equation}
Upon a hopping expansion of $\log M$, $Tr \log M$ represents a sum of 
single loops with all sizes and shapes. The determinant is then 
the sum of all multiple loops. The fermion loops can be separated into
two classes. One is those which do not go across the time boundary and
represent virtual quark-antiquark pairs; the other includes those
which wraps around the time boundary which represent external quarks
and antiquarks. The configuration with a baryon number one which entails three
quark loops wrapping around the time boundary will have
an energy $M_B$ higher than that with zero baryon number. Thus, it is
weighted with the probability $e^{- M_B N_t a_t}$ compared with the one 
with no net baryons. We see from the above discussion that the fermion 
determinant contains a superposition of sectors of all baryon numbers, positive, 
negative and zero. At zero temperature where $M_B N_t a_t \gg 1$, the 
zero baryon sector dominates and all the other baryon sectors are exponentially 
suppressed. It is obvious that to avoid the overlap problem, one
needs to select a definite nonzero baryon number sector and stay in it
throughout the Markov chain of updating configurations. To select a
particular baryon sector from the determinant can be achieved by
the following procedure~\cite{fab95}: first, assign an $U(1)$ phase factor 
$e^{-i \phi}$ to the links between the time slices $t$ and $t + 1$ 
so that the link $U/U^{\dagger}$ is multiplied by $e^{-i \phi}/e^{i \phi}$;
then the particle number projection can be carried out through the
Fourier transformation of the fermion determinant like in the BCS theory
\begin{equation}  \label{projection}
P_N = \frac{1}{2 \pi} \int_0^{2 \pi} d\phi e^{-i \phi N} det M[\phi]
\end{equation}
where $N$ is the net particle number, i.e. particle minus antiparticle.
Note that all the virtual quark loops which do not reach the time
boundary will have a net phase factor of unity; only those with a 
net N quark loops across the time boundary will have a phase factor
$e^{i \phi N}$ which can contribute to the integral in Eq. (\ref{projection}).
Since QCD in canonical formulation does not break $Z(3)$
symmetry, it is essential to take care that the ensemble is
canonical with respect to triality. To this end, we shall consider
the triality projection~\cite{fab95,fbm95} to the zero triality sector
\begin{equation}
det_0 M = \frac{1}{3} \sum_{k = 0, \pm 1} det M [\phi + k 2\pi/3].
\end{equation}   
This amounts to limiting the quark number N to a multiple of 3. Thus
the triality zero sector corresponds to baryon sectors with integral
baryon numbers.

   Another essential ingredient to circumvent the overlap problem is
to stay in the chosen nonzero baryon sector so as to avoid mixing
with the zero baryon sector with exponentially large weight.
This can be achieved by preforming the baryon number projection
as described above {\it before} the accept/reject step in the 
Monte Carlo updating of the gauge configuration. If this is not
done, the accepted gauge configuration will be biased toward
the zero baryon sector and it is very difficult to project out
the nonzero baryon sector afterwords. This is analogous to the
situation in the nuclear many-body theory where it is known~\cite{wong75} that the
variation after projection (Zeh-Rouhaninejad-Yoccoz method~\cite{zeh65,ry66}) 
is superior than the variation before projection (Peierls-Yoccoz method~\cite{py57}).
The former gives the correct nuclear mass in the 
case of translation and yields much improved wave functions in mildly deformed 
nuclei than the latter.

\begin{figure}[tb]
\includegraphics{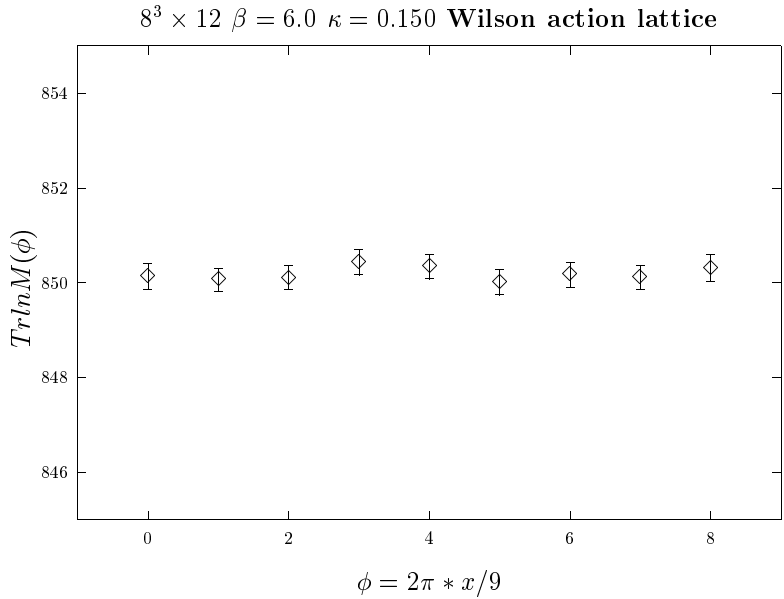}
\vspace{8cm}
\caption{$Tr \log M[\phi]$ for a $8^3 \times 12$ configuration with Wilson
action as a function of $\phi$. }
\end{figure}

\begin{figure}[tb]
\includegraphics{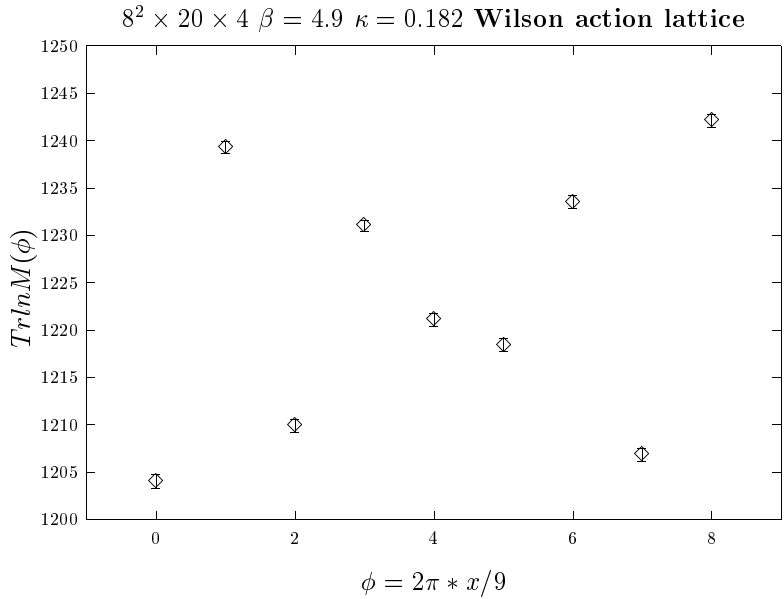}
\vspace{8cm}
\caption{$Tr \log M[\phi]$ for a $8\times 20^2 \times 4$ finite temperature 
configuration with dynamical fermion.}
\end{figure}

   To illustrate the overlap problem, we plot in Fig.1 \,$Tr \log M[\phi]$ for
a configuration of the $8^3 \times 12$ lattice with the Wilson action
with $\beta = 6.0$ and $\kappa = 0.150$ which is obtained with 500 $Z_2$
noises. We see that the it is rather flat in $\phi$ indicating that
the Fourier transform in Eq. (\ref{projection}) will mainly favor the
zero baryon sector. On the other hand, at finite temperature, it is
relatively easier for the quarks to be excited so that the zero baryon
sector does not necessarily dominate other baryon sectors. Another way of seeing this
is that the relative weighting factor $e^{- M_B N_t a_t}$ can be
$O(1)$ at finite temperature. Thus, it should be easier to project out
the nonzero baryon sector from the determinant. We plot in Fig. 2
a similarly obtained $Tr \log M[\phi]$ for a configuration of the 
$8 \times 20^2 \times 4$ lattice with $\beta = 4.9$ and $\kappa = 0.182$.
We see from the figure that there is quite a bit of wiggling in this
case as compared to that in Fig. 1 indicating that it is easier to
project out a nonzero baryon sector through the Fourier transform at finite
temperature.

   We should mention that while we think we can overcome the 
overlap problem and the determinant $det M[\phi]$ is real in this approach, 
nevertheless in view of the fact that the Fourier transform in 
Eq. (\ref{projection}) involves the quark number $N$ 
the canonical approach may still have the sign problem
at the thermodynamic limit when $N$ and $V$ are very large. 
However,  we think it might work for small $N$ 
such as 3 or 6 for one or two baryons in a finite $V$. This should
be a reasonable start for practical purposes.  

   While it is clear what the algorithm in the canonical approach
entails, there are additional practical requirements for the
algorithm to work. These include an unbiased estimation of the
huge determinant in lattice QCD and, moreover, a Monte Carlo 
algorithm which accommodates the unbiased estimate of the
probability. We shall discuss them in the following sections.

\section{A Noisy Monte Carlo Algorithm}

There are problems in physics which involve extensive quantities such 
as the fermion determinant which require $V^3$ steps to compute exactly.
Problems of this kind with large volumes are not numerically applicable
with the usual Monte Carlo algorithm which require an exact evaluation of 
the probability ratios in the accept/reject step. 
To address this problem, Kennedy and Kuti \cite{kk85} proposed a Monte 
Carlo algorithm which admits stochastically estimated transition probabilities 
as long as they are unbiased. But there is a drawback. 
The probability could lie outside the interval between $0$
and $1$ since it is estimated stochastically. This  
probability bound violation will destroy detailed balance
and lead to systematic bias. To control the probability violation
with a large noise ensemble can be costly.

We propose a noisy Monte Carlo algorithm which avoids this difficulty
with two Metropolis accept/reject steps.
Let us consider a model with Hamiltonian $H(U)$ where $U$ collectively 
denotes the dynamical variables of the system.
The major ingredient of the new approach is
to transform the noise for the stochastic estimator into stochastic variables.
The partition function of the model can be written as
\begin{eqnarray}  \label{Z}
Z &=&  \int [DU]\, e^{-H(U)} \nonumber \\ 
  &=& \int [DU][D\xi]P_\xi(\xi)\, f(U,\xi).
\end{eqnarray}
where $f(U,\xi)$ is an unbiased estimator of $e^{-H(U)}$ from the 
stochastic variable $\xi$ and $P_\xi$ is the probability distribution 
for $\xi$.

The next step is to address the lower probability-bound violation. One
first notes that one can write the expectation value
of the observable $O$ as
\begin{equation}  \label{O}
\langle O \rangle = \int[DU][D\xi]\,P_\xi(\xi)
  \,O(U)\,{\rm sign}(f)\,|f(U,\xi)|/Z,
\end{equation}
where $sign(f)$ is the sign of the function $f$.
After redefining the partition function to be
\begin{equation}
 Z = \int [DU][D\xi]P_\xi(\xi)\, |f(U,\xi)|,
\end{equation}
which is semi-positive definite, 
the expectation of $O$ in Eq. (\ref{O}) can be rewritten as
\begin{equation}  \label{Onew}
\langle O \rangle = \langle O(U) \,{\rm sign}(f) \rangle/\langle 
 {\rm sign}(f) \rangle.
\end{equation} 
As we see, the sign of $f(U,\xi)$ is not a part of the probability any more
but a part in the observable.
Notice that this reinterpretation is possible because the sign of
$f(U,\xi)$ is a state function which depends on the configuration of $U$ and
$\xi$. 

It is clear then, to avoid the problem of lower probability-bound 
violation, the accept/reject criterion has to be factorizable into
a ratio of the new and old probabilities so that the sign of the
estimated $f(U,\xi)$ can be absorbed into the observable.  
This leads us to the Metropolis accept/reject criterion which
incidentally cures the problem of upper probability-bound violation at
the same time. It turns out two accept/reject steps are needed in general.
The first one is to propose updating of $U$ via some procedure
while keeping the stochastic variables $\xi$ fixed.
The acceptance probability $P_a$ is
\begin{equation}  \label{met1}
 P_a(U_1,\xi \rightarrow U_2,\xi)\,\,= \,\,
  {\rm min}\Bigl(1, \frac{|f(U_2,\xi)|}
                    {|f(U_1,\xi)|}\Bigr)\,\,  .
\end{equation}
The second accept/reject step involves the refreshing of the stochastic 
variables $\xi$ according to the probability distribution $P_{\xi}(\xi)$ while 
keeping $U$ fixed. The acceptance probability is 
\begin{equation}   \label{met2}
  P_a(U,\xi_1\rightarrow U,\xi_2)\,\,= \,\,
  {\rm min}\Bigl(1,\frac{|f(U,\xi_2)|}{|f(U,\xi_1)|}\Bigr)\,\,  .
\end{equation}
It is obvious that there is neither lower nor upper probability-bound 
violation in either of these two Metropolis accept/reject steps. 
Furthermore, it involves the ratios of separate state functions so that
the sign of the stochastically estimated probability $f(U,\xi)$ can be 
absorbed into the observable as in Eq. (\ref{Onew}).  
 
  Detailed balance can be proven to be satisfied and it is unbiased
~\cite{lls99}. Therefore, this is an exact algorithm.

\section{Noisy Monte Carlo with Fermion Determinant}

One immediate application of NMC is lattice QCD with dynamical fermions.
The action is composed of two parts -- the pure gauge action
$S_g(U)$ and a fermion action $S_F(U) = - Tr \ln M(U)$.
Both are functionals of the gauge link variables $U$. 

To find out the explicit form of $f(U,\xi)$, we
note that the fermion determinant can be calculated stochastically as
a random walk process~\cite{bk85}
\begin{equation}
e^{Tr\ln M} = 1 + Tr\ln M (1 + \frac{Tr \ln M}{2} (1 + \frac{Tr \ln M}{3}
(...))) \,\, .
\end{equation} 
This can be expressed in the following integral
\begin{eqnarray}  \label{trln}
&& e^{Tr\ln M}=\int \prod_{i =1}^{\infty}  d\,\eta_i \,
P_{\eta}(\eta_i)
     \int_{0}^1 \prod_{n =2}^{\infty}  d\,\rho_n  \nonumber \\
&& [1 + \eta_1^{\dagger} \ln M \eta_1
  (1 + \theta(\rho_2 - \frac{1}{ 2}) \eta_2^{\dagger} \ln M \eta_2
  (1 + \theta(\rho_3 - \frac{2}{ 3}) \eta_3^{\dagger} \ln M \eta_3 (...],
\end{eqnarray}
where $P_{\eta}(\eta_i)$ is the probability distribution for the
stochastic variable $\eta_i$. It can be the Gaussian noise or the $Z_2$ noise 
($P_{\eta}(\eta_i) = \delta(|\eta_i| -1)$ in this case). The latter is
preferred since it has the minimum variance~\cite{dl94}. $\rho_n$ is
a stochastic variable with uniform distribution between 0 and 1.
This sequence terminates stochastically in finite time and only the
seeds from the pseudo-random number generator need to be stored in practice.
The function $f(U,\eta,\rho)$ ( 
$\xi$ in Eq. (\ref{Z}) is represented by two stochastic variables 
$\eta$ and $\rho$ here) is represented by the part of the integrand 
between the the square brackets in Eq. (\ref{trln}). 
One can then use the efficient 
Pad\'{e}-Z$_2$ algorithm~\cite{TDLY} to calculate the $\eta_i\ln M \eta_i$ in
Eq. (\ref{trln}). We shall discuss this in the next section.

Finally, there is a practical concern that $Tr\ln M$ can be large
so that it takes a large statistics to have a reliable estimate of
$e^{Tr\ln M}$ from the series expansion in Eq. (\ref{trln}). In general, 
for the Taylor expansion $e^x = \sum x^n/n!$, the series will start to
converge when $x^n/n! > x^{n + 1}/(n + 1)!$. This happens at $n = x$.
For the case $x = 100$, this implies that one needs to have more than 100! 
stochastic configurations in the Monte Carlo
integration in Eq. (\ref{trln}) in order to have a convergent estimate.
Even then, the error bar will be very large. To avoid this difficulty, one
can implement the following strategy. First, one notes that since the 
Metropolis accept/reject involves the ratio of exponentials, one can subtract 
a universal number $x_0$ from the exponent $x$ in the Taylor expansion 
without affecting the ratio. Second, one can use a specific form of the
exponential
to diminish the value of the exponent. In other words, one can replace $e^x$ 
with $(e^{(x - x_0)/N})^N$ to satisfy $|x - x_0|/N < 1$. The best choice for
$x_0$ is $\overline{x}$, the mean of $x$. In this case, the variance of
$e^x$ becomes $e^{\delta^2/N} -1$. 

\section{The Pad\'e -- Z$_2$ Method of Estimating Determinants }
Now we shall discuss a very efficient way of estimating the fermion
determinant stochastically~\cite{TDLY}.  

\subsection{Pad\'e approximation}
The starting point for the method is the
Pad\'e approximation of the logarithm function.
The Pad\'e approximant to $\log (z)$  of order $[K,K]$ at $z_0$
is a rational function $N(z)/D(z)$ where deg~$N(z)$~=~deg~$D(z)$ = $K$,
whose value and first $2K$ derivatives agree with $\log z$ at the
specified point $z_0$. When the Pad\'e approximant $N(z)/D(z)$
is expressed in partial fractions, we obtain
\begin{equation}
\log z \approx b_0 + \sum_{k=1}^{K}\left( \frac{b_k}{z +c_k} \right),
\label{Pade_apprx}
\end{equation}
whence it follows
\begin{equation}    \label{Tr_apprx}
\mbox {log det }{\bf M} =
\mbox {Tr log}{\bf M} \approx b_0 Tr {\bf I} +
               \sum_{k=1}^K  b_k\cdot Tr ({\bf M} +c_k{\bf I})^{-1}.
\end{equation}

   The Pad\'e approximation is not limited to the real axis. As long as 
the function is in the analytic domain, i. e. away from the
cut of the log, say along the negative real axis, the Pad\'e approximation
can be made arbitrarily accurate by going to a higher order $[K,K]$ and
a judicious expansion point to cover the eigenvalue domain of the problem.

  \subsection{Complex Z$_2$ noise trace estimation}
Exact computation of the trace inverse for $N \times N$ matrices is very
time consuming  for matrices
of size $N \sim 10^6$.  However, the complex Z$_2$ noise method
has been shown to provide an efficient stochastic estimation of the
trace~\cite{dl94,dll95,Eick96}. In fact, it has been proved to be an
optimal choice for the noise, producing a {\it minimum}
variance~\cite{bmt94}.
 
The complex Z$_2$ noise estimator can be briefly described as follows
\cite{dl94,bmt94}.
We construct L noise vectors
$\eta^1,\eta^2,\cdots, \eta^L$ where
$\eta^j = \{ \eta^j_1, \eta^j_2, \eta^j_3, \cdots, \eta^j_N \}^{T}$,
as follows. Each element $\eta^j_n$ takes one of the four values
$\{\pm 1 ,\,\pm \imath \}$  chosen independently with equal probability.
It follows from the statistics of $\eta^j_n$ that
\begin{eqnarray}
E[<\eta_n>] \equiv E[\frac {1}{L} \sum_{j=1}^L \eta^j_n]=0,~~~~~~~~
E[<\eta_m^{\star} \eta_n>] \equiv
        E[\frac {1}{L} \sum_{j=1}^L \eta_m^{\star
j}\eta_n^j]=\delta_{mn}.~~~~~~~~~~~~~
\end{eqnarray}
The vectors can be used to construct an unbiased estimator for the trace
inverse of a given matrix $M$  as follows:
\begin{eqnarray*}
E[<\eta^\dagger  \mM^{-1} \eta>]&\equiv&
E[\frac 1L \sum_{j=1}^L \sum_{m,n=1}^N \eta^{\star j}_m M^{-1}_{m,n}
\eta^j_n]
~~~~~~~~~~~~~~~~~~~~~~~~~~~~~~~~~~~~~~~~~~~~~~~\\
&=& \sum_n^N M^{-1}_{n,n}  +
(\sum_{m\neq n}^N M^{-1}_{m,n})[\frac 1L \sum_j^L \eta^{\star j}_m
\eta^j_n]\\
=&& \mbox {Tr }\mM^{-1}.
\end{eqnarray*}
The variance of the estimator is shown to be \cite{bmt94}
\begin{eqnarray*}
\sigma^2_M &\equiv& \Var[<\eta^\dagger \mM^{-1}\eta>]=
E\left[ |<\eta^\dagger \mM^{-1}\eta>-\mbox {Tr }\mM^{-1}|^2\right]
~~~~~~~~~~~~~~~~~~~~~~~~~~~~\\
 &=& \frac 1L \sum_{m\neq n}^N M^{-1}_{m,n}(M^{-1}_{m,n})^\star
      = \frac 1L \sum_{m\neq n}^N |M^{-1}_{m,n}|^2~.
\end{eqnarray*}
 
The stochastic error of the complex Z$_2$ noise estimate results only
from the off-diagonal
entries of the inverse matrix (the same is true for Z$_n$ noise for
any n). However, other noises (such as Gaussian)
have additional errors arising from diagonal entries. This is why the
Z$_2$ noise has minimum variance. For example, it has been demonstrated
on a $16^3 \times 24$ lattice with $\beta = 6.0$ and $\kappa = 0.148$ for
the Wilson action that the Z$_2$ noise standard deviation is smaller
than that of the Gaussian noise by a factor of 1.54~\cite{dl94}.

Applying the complex Z$_2$ estimator to the expression for the
$Tr log {\bf M}$ in Eq. (\ref{Tr_apprx}), we find
\begin{eqnarray}
&&\sum_{k} b_k  \mbox {Tr}(M+c_k)^{-1} \nonumber \\ 
&&\approx \frac 1L \sum_{k}^K \sum_j^{L} b_k \eta^{j \dagger}
  (M+c_k)^{-1} \eta^j  \nonumber\\
&&=  \frac {1}{L} \sum_{j}^L \sum_{k=1}^{K} b_k \eta^{j \dagger} \xi^{k,j},
\end{eqnarray}
where $\xi^{k,j}= (M+c_k {\bf I})^{-1} \eta^j$ are the solutions of
\begin{equation}   \label{col_inv1}
(M + c_k {\bf I} )\xi^{k,j} = \eta^j, 
\end{equation}
Since $M + c_k {\bf I}$ are shifted matrices with constant diagonal
matrix elements,
Eq. (\ref{col_inv1})  can be solved collectively
for all values of $c_k$
within one iterative process by several algorithms, including
the Quasi-Minimum Residual (QMR)
~\cite{fng95}, Multiple-Mass Minimum Residual (M$^3$ R)~\cite{ggl96},
and GMRES\cite{fg96}. We have adopted the M$^3$ R algorithm,
which has been shown to be about 2 times faster than the conjugate
gradient algorithm, and the overhead for the multiple $c_k$ is only
~8\%~\cite{ydl96}. The only price to pay is memory: for each $c_k$,
a vector of the solution needs to be stored.
Furthermore, one observes that $c_k > 0$. This improves the
conditioning of $({\bf M} + c_k {\bf I})$ since the eigenvalues of
${\bf M}$ have positive real parts.  Hence, we expect faster convergence for
column inversions for Eq. (\ref{col_inv1}).
 
 In the next section,
we describe a method which significantly reduces the stochastic error.
 
\subsection{Improved PZ estimation with unbiased subtraction}

In order to reduce the variance of the estimate, we introduce a
suitably chosen set of traceless
$N \times N$ matrices $\mQ^{(p)}$, i.e. which satisfy
$\sum_{n=1}^{N} \mQ^{(p)}_{n,n} = 0,\, p = 1 \cdots P$.
The expected value and variance for  the modified estimator
$<\eta^\dagger(\mM^{-1} - \sum_{p=1}^P \gl_p \mQ^{(p)}) \eta>$
are given by
\begin{eqnarray}
E[<\eta^\dagger (\mM^{-1} - \sum_{p=1}^P \gl_p \mQ^{(p)})) \eta>]&=&
 \mbox {Tr }\mM^{-1}~, ~~~~~~~~~~~~~~~~~~~~~~~~~~~~~~~~~\\
\Delta_M(\gl) = \Var[<\eta^\dagger (\mM^{-1} -
                      \sum_{p=1}^P \gl_p \mQ^{(p)}) \eta>]
 &=& \frac 1L \sum_{m\neq n} |\mM^{-1}_{m,n}-
    \sum_{p=1}^P \gl_p \mQ^{(p)}_{m,n})|^2 ~,  \label{red_var}
\end{eqnarray}
for any values of the real parameters $\gl_p$. In other words,
introducing the matrices $\mQ^{(p)}$  into the estimator produces no
bias, but
may reduce the error bars if the $\mQ^{(p)}$ are chosen judiciously.
Further, $\gl_p$ may be varied at will to achieve a minimum
variance estimate:  this corresponds to a least-squares fit to the
function
$\eta^\dagger \mM^{-1} \eta$ sampled at points $\eta_j,~j=1 \cdots L$,
using the fitting
functions $\left\{1, \eta^\dagger \mQ^{(p)} \eta \right\},~
p=1 \cdots P$.

We now turn to the question of choosing suitable
traceless matrices $\mQ^{(p)}$  to use in the modified estimator.
One possibility for the Wilson fermion matrix $\mM = \mI - \kappa\mD$
is suggested by the hopping parameter --- $\kappa$ expansion of the
inverse matrix,
\begin{eqnarray}
(\mM + c_k \mI )^{-1} &=& \frac {1}{\mM + c_k \mI }
= \frac {1}{(1+c_k)(\mI - \frac {\kappa}{(1+c_k)} \mD) }\nonumber
   ~~~~~~~~~~~~~~~~~~~~~~~~\\
&=& \frac {\mI}{1+c_k} + \frac {\kappa}{(1+c_k)^2} \mD
+\frac {\kappa^2}{(1+c_k)^3} \mD^2
+\frac {\kappa^3}{(1+c_k)^4} \mD^3 + \cdots~.
\end{eqnarray}
This suggests choosing the matrices $\mQ^{(p)}$
from among those matrices in the hopping parameter expansion which
are traceless:
\begin{eqnarray*}
\mQ^{(1)} &=&\frac {\kappa}{(1+c_k)^2}\mD,~~~~~~~~~~~~~
                  ~~~~~~~~~~~~~~~~~~~~~\\
\mQ^{(2)} &=&\frac {\kappa^2}{(1+c_k)^3}\mD^2, \\
\mQ^{(3)} &=&\frac {\kappa^3}{(1+c_k)^4}\mD^3, \\
\mQ^{(4)} &=&\frac {\kappa^4}{(1+c_k)^5}(\mD^4-{\rm Tr}\mD^4), \\
\mQ^{(5)} &=&\frac {\kappa^5}{(1+c_k)^6}\mD^5, \\
\mQ^{(6)} &=&\frac {\kappa^6}{(1+c_k)^7}(\mD^6-{\rm Tr}\mD^6), \\
\mQ^{(2r+1)} &=& \frac {\kappa^{2r+1} }{(1+c_k)^{2r+2}}\mD^{2r+1},
~~~~~                          r=3,4,5, \cdots~.
\end{eqnarray*}
It may be verified that all of these matrices are traceless.
In principle, one can include all the even powers which entails the
explicit calculation of all the allowed loops in $Tr D^{2r}$. In this
manuscript we have only included $\mQ^{(4)}$, $\mQ^{(6)}$,
and $\mQ^{(2r+1)}$.

\subsection{Computation of $Tr\, log\,M$}
Our numerical computations were carried out with the Wilson action
on the $8^3 \times 12$ ($N$ = 73728) lattice with $\beta =5.6$.
We use the HMC with pseudofermions to generate gauge configurations.
With a cold start, we obtain the fermion matrix $\mM_1$ after the
plaquette becomes stable. The trajectories are traced with $\tau = 0.01$
and 30 molecular dynamics steps using $\kappa = 0.150$.
$\mM_2$ is then obtained from $\mM_1$ by an accepted trajectory run.
Hence $\mM_1$ and $\mM_2$ differ by a continuum
perturbation, and $\log [\det \mM_1/\det \mM_2 ] \sim {\cal O}(1)$.
 
  We first calculate $\log \det\mM_1$ with different orders of Pad\'e
expansion around $z_0 = 0.1$ and $z_0 = 1.0$. We see from Table 1
that the 5th order Pad\'e does not give the same answer for two
different expansion points, suggesting that its accuracy is not
sufficient for the range of eigenvalues of $\mM_1$. Whereas, the 11th
order Pad\'e gives the same answer within errors.
Thus, we shall choose P[11,11]$(z)$ with $z_0=0.1$ to perform the
calculations from this point on.

\begin{table} \begin{center}
\caption{ Unimproved and improved PZ estimates for
log [det{\bf M}$_1$]  with 100 complex 
Z$_2$ noise vectors. $\kappa = 0.150$.
 }
\bigskip
\bigskip
\bigskip
{\begin{tabular}{c c c c c c } \hline\hline
$P[K,K](z)$ &  $K=$  &  5  &   7   &  9   &  11     \\ \hline
$z_0 = 0.1$ & Original: & 473(10) & 774(10) & 796(10) & 798(10) \\
~ & Improved: & 487.25(62) & 788.17(62) & 810.83(62) & 812.33(62) \\ \hline
$z_0 = 1.0$ &Original: & 798(10) & 798(10) & 798(10) & 799(10)\\
~ & Improved: & 812.60(62) & 812.37(62) & 812.36(62) & 812.37(62)\\
\hline\hline
\end{tabular}}
\end{center} 
\end{table}

\begin{table} 
\begin{center}
\caption{ Central values for improved stochastic estimation of
log[det {\bf M}$_1$] and $r$th--order improved Jackknife errors
$\delta_r$ are given for different numbers of Z$_2$ noise vectors.
$\kappa$ is 0.150 in this case.
}
\bigskip
\bigskip
\bigskip
{\begin{tabular}{ c c c c c c c c c c} \hline\hline
 \# Z$_2$    & 50 & 100 & 200 & 400 & 600 & 800 & 1000 & 3000 & 10000 \\ \hline
 0$^{th}$    &  802.98   &  797.98   &  810.97   &  816.78   &  815.89
             &  813.10   &  816.53   &  813.15   &  812.81    \\
 $\delta_0$  & $\pm$14.0 & $\pm$9.81 & $\pm$7.95 & $\pm$5.54 & $\pm$4.47
	     & $\pm$3.83 & $\pm$3.41 & $\pm$1.97 & $\pm$1.08  \\ \hline
  1$^{st}$   &  807.89   &  811.21   &  814.13   &  815.11   &  814.01
             &  814.62   &  814.97   & --- & ---  \\
 $\delta_1$  & $\pm$4.65 & $\pm$3.28 & $\pm$2.48 & $\pm$1.84 & $\pm$1.50
             & $\pm$1.29 & $\pm$1.12 &  -  & -    \\ \hline
  2$^{nd}$   &  813.03   &  812.50   &  811.99   &  812.86   &  811.87
             &  812.89   &  813.04 & ---  & ---   \\
 $\delta_2$  & $\pm$2.46 & $\pm$1.65 & $\pm$1.34 & $\pm$1.01 & $\pm$0.83
             & $\pm$0.72 & $\pm$0.64 & -   & -    \\ \hline
  3$^{rd}$   &  812.07   &  812.01   &  811.79   &  812.44   &  812.18
             &  812.99   &  813.03 & ---  & ---   \\
 $\delta_3$  & $\pm$1.88 & $\pm$1.31 & $\pm$1.01 & $\pm$0.74 & $\pm$0.58
             & $\pm$0.51 & $\pm$0.44 & -  & -    \\ \hline
  4$^{th}$   &  812.28   &  812.52   &  812.57   &  812.86   &  812.85
             &  813.25   &  813.40 & ---   & ---  \\
 $\delta_4$  & $\pm$1.20 & $\pm$0.94 & $\pm$0.68 & $\pm$0.48 & $\pm$0.39
             & $\pm$0.35 & $\pm$0.30 & -  & -    \\ \hline
  5$^{th}$   &  813.50   &  813.07   &  813.36   &  813.70   &  813.47
             &  813.54   &  813.50 & ---  & ---  \\
 $\delta_5$  & $\pm$0.82 & $\pm$0.62 & $\pm$0.47 & $\pm$0.34 & $\pm$0.29
             & $\pm$0.25 & $\pm$0.22 & -  & -    \\ \hline
  6$^{ts}$   &  813.54   &  813.23   &  813.22   &  813.28   &  813.20
	     &  813.37   &  813.26 & ---  & ---  \\
 $\delta_6$  & $\pm$0.67 & $\pm$0.49 & $\pm$0.35 & $\pm$0.25 & $\pm$0.21
             & $\pm$0.18 & $\pm$0.16 & -  & -    \\ \hline
  7$^{ts}$   &  814.18   &  813.74   &  813.44   &  813.42   &  813.39
             &    ---    &    ---  & ---  & ---  \\
 $\delta_7$  & $\pm$0.44 & $\pm$0.36 & $\pm$0.26 & $\pm$0.19 & $\pm$0.16
             &     -     &     -     & -  & -    \\ \hline
  9$^{th}$   &  813.77   &  813.62   &  813.49   &  813.40   &  813.43
             &    ---    &    ---  & ---  & ---  \\
 $\delta_9$  & $\pm$0.40 & $\pm$0.30 & $\pm$0.22 & $\pm$0.16 & $\pm$0.14
	     &     -     &     -     & -  & -    \\ \hline
 11$^{th}$   &  813.54   &  813.41   &  813.45   &  813.44   &  813.43
             &    ---    &    ---  & ---  & ---  \\
$\delta_{11}$& $\pm$0.38 & $\pm$0.27 & $\pm$0.21 & $\pm$0.15 & $\pm$0.13
             &     -     &     -     & - & -    \\ \hline\hline
\end{tabular}}
\end{center} 
\end{table}

\begin{figure}[ht]
\includegraphics{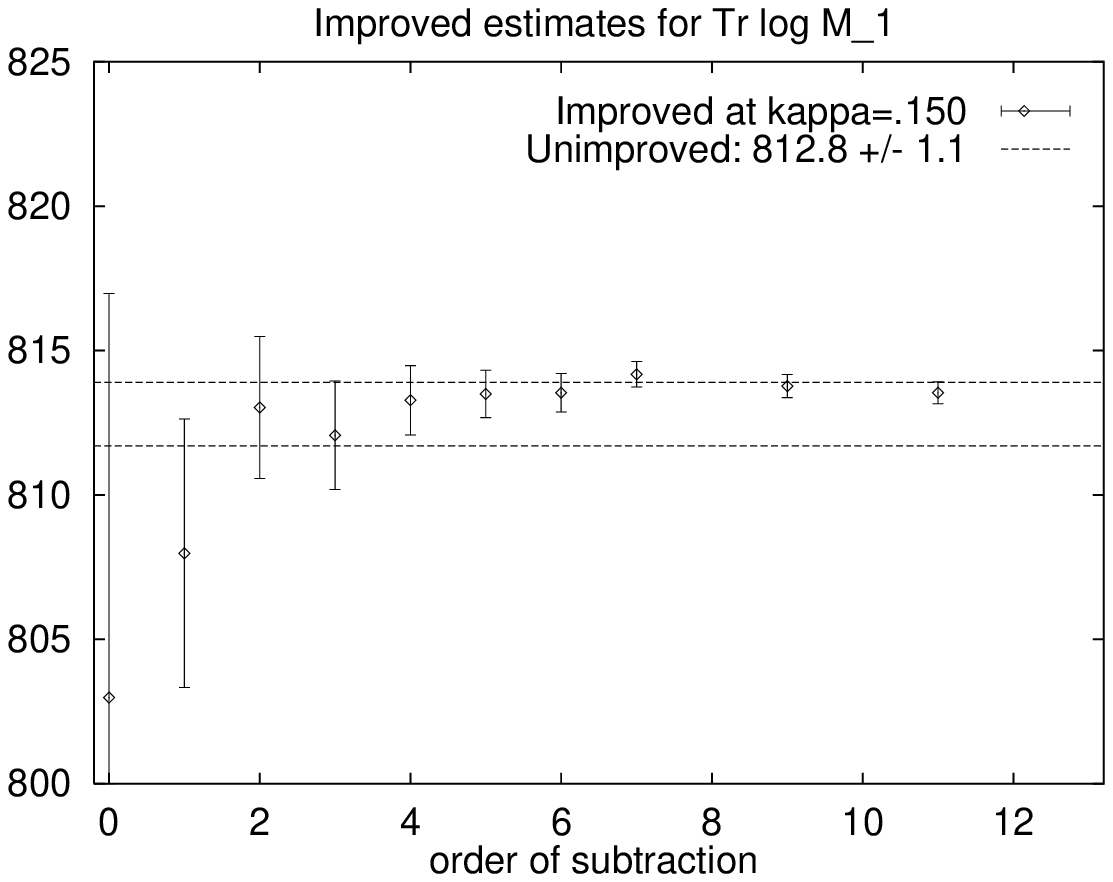}
\caption{The improved PZ estimate of $\Tr \log \mM_1$ with 50 noises
as a function
of the order of subtraction and compared to that of unimproved estimate
with 10,000 noises. The dashed lines are drawn with a distance of
1 $\sigma$ away from the central value of the unimproved estimate.}
\end{figure}

In Table 2, we give the results of improved estimations for
$\Tr \log \mM_1$.
We see that the variational technique
described above can reduce the data fluctuations by more than an order of
magnitude. For example, the unimproved error $\delta_0 = 5.54$ in
Table 2 for
400 Z$_2$ noises is reduced to $\delta_{11} = 0.15$ for the subtraction
which
includes up to the $\mQ^{11}$ matrix. This is 37 times smaller. Comparing
the central values in the last row (i.e. the $11^{th}$ order improved)
with that of unimproved estimate with 10,000 Z$_2$ noises, we see that
they are the same within errors. This verifies that the variational
subtraction scheme that we employed does not introduce biased
errors. The improved estimates of $\Tr \log \mM_1$ from 50
Z$_2$ noises with
errors $\delta_{r}$ from Table 2 are plotted in comparison with the central
value of the unimproved estimate from 10,000 noises in Fig. 3.

\section{Implementation of the Kentucky Noisy Monte Carlo Algorithm}

We have recently implemented the above Noisy Monte Carlo algorithm to
the simulation of lattice QCD with dynamical fermions by incorporating
the full determinant directly~\cite{jhl02}. Our algorithm uses pure gauge field
updating with a shifted gauge coupling to minimize fluctuations in the 
trace log is the Wilson Dirac matrix. It gives the
correct results as compared to the standard Hybrid Monte Carlo simulation.
However, the present simulation has a low acceptance rate due to the 
pure gauge update and results in long autocorrelations. We are in the
process of working out an alternative updating scheme with molecular dynamics
trajectory to include the feedback of the determinantal effects on
the gauge field which should be more efficient than the pure gauge
update.

\section{Summary}
After reviewing the finite density algorithm for QCD with the chemical
potential, we propose a canonical approach by projecting out the
definite baryon number sector from the fermion determinant and stay in
the sector throughout the Monte Carlo updating. This should circumvent the overlap
problem. In order to make the algorithm practical, one needs an
efficient way to estimate the huge fermion determinant and a
Monte Carlo algorithm which admits unbiased estimates of the probability
without upper unitarity bound violations. These are achieved with the
Pad\'{e}-Z$_2$ estimate of the determinant and the Noisy Monte Carlo
algorithm. So far, we have implemented the Kentucky Noisy Monte Carlo
algorithm to incorporate dynamical fermions in QCD on a 
relatively small lattice and medium heavy quark based on pure gauge updating.
As a next step, we will work on a more efficient updating algorithm and
project out the baryon sector to see if the finite density algorithm proposed
here will live up to its promise.

 \section*{Acknowledgments}

This work is partially supported by the U.S. DOE grant DE-FG05-84ER40154. The 
author wishes to thank M. Faber for introducing the subject of the finite density
to him and Shao-Jing Dong for providing some unpublished figures.
Fruitful discussions with M. Alford, I.Barbour, U.-J. Wiese, and R. Sugar are
acknowledged. He would also thank the organizer, Ge Mo-lin for
the invitation to attend the conference and his hospitality.

\end{document}